\documentclass[ preprintnumbers,amsmath,amssymb,floatfix]{revtex4}

\usepackage{graphicx}
\usepackage{dcolumn}
\usepackage{bm}

\newcommand{\beq}{\begin{equation}}
\newcommand{\eeq}{\end{equation}}
\newcommand{\bey}{\begin{eqnarray}}
\newcommand{\eey}{\end{eqnarray}}

\begin{document}

\title{Macroscopic wormholes in noncommutative geometry}

\author{Peter K.F. Kuhfittig}
\email{kuhfitti@msoe.edu} \affiliation{Department of Mathematics,
Milwaukee School of Engineering, Milwaukee, Wisconsin 53202-3109,
USA}

\begin{abstract}\noindent
The purpose of this paper is to show that wormholes in
noncommutative geometry can be macroscopic, based in part on
an earlier study.  The necessary violation of the weak
energy condition is attributable to the
noncommutative-geometry background rather than to the use
of ``exotic matter."  The zero tidal forces make the
wormhole suitable for humanoid travelers.  Also discussed
is the stability to linearized radial perturbations.
\end{abstract}

\maketitle
\noindent
\text{PACS numbers: 04.20.Jb, 04.20.Gz, 11.10.Nx}

\section{Introduction}\noindent
An important outcome of string theory is the realization
that coordinates may become noncommutative operators on a
$D$-brane \cite{eW96, SW99}.  The result is a fundamental
discretization of spacetime due to the commutator
$[\textbf{x}^{\mu},\textbf{x}^{\nu}]=i\theta^{\mu\nu}$, where
$\theta^{\mu\nu}$ is an antisymmetric matrix, similar to the
way that the Planck constant $\hbar$ discretizes phase space.
Noncommutativity replaces point-like objects by smeared
objects \cite{SS03, NSS06, SSN09}.  The goal is to eliminate
the divergences that normally appear in general relativity.
The smearing effect can be accomplished by using a Gaussian
distribution of minimal length $\sqrt{\theta}$ instead of
the Dirac delta function in the following manner:  the
energy-density of the static and spherically symmetric
and particle-like gravitational source has the form
\cite{NSS06, mR11}
\begin{equation}\label{E:rho}
  \rho(r)=\frac{M}{(4\pi\theta)^{3/2}}e^{-r^2/4\theta}.
\end{equation}
Here the mass $M$ is diffused throughout the region
of linear dimension $\sqrt{\theta}$ due to the uncertainty.
The noncommutative geometry is an intrinsic property of
spacetime and does not depend on particular features such
as curvature.

Noncommutative geometry has proved to be an effective
tool in several areas.  Thus Garattini and
Lobo \cite{GL09} obtained a self-sustained wormhole
using a semiclassical analysis.   Kuhfittig
\cite{pK12} showed that a special class of thin-shell
wormholes posses small regions of stability around the
thin shell even though they are unstable in classical
general relativity. Rahaman et al. \cite{fR12}
concluded that a noncommutative-geometry background
is sufficient for producing stable circular orbits
in a typical galaxy without the need for dark matter.

Turning our attention now to traversable wormholes,
recall that the line element describing a
Morris-Thorne wormhole is given by \cite{MT88}

\begin{equation}
    ds^2=-e^{\Phi(r)}dt^2+\frac{dr^2}{1-b(r)/r}
    +r^2(d\theta^2+\text{sin}^2\theta\,d\phi^2).
\end{equation}
Here $b=b(r)$ is called the \emph{shape function} and
$\Phi=\Phi(r)$ the \emph{redshift function}, which must
be everywhere finite to prevent an event horizon.  For
the shape function we must have $b(r_0)=r_0$, where
$r=r_0$ is the radius of the \emph{throat} of the
wormhole.  In addition, $b'(r_0)<1$ and $b(r)<r$
to satisfy the \emph{flare-out condition} \cite{MT88}.
The flare-out condition, in turn, can only be
satisfied by violating the \emph{weak energy condition},
to be discussed below.  For ordinary matter, this
violation can only be accomplished by the use of
``exotic matter."

The parameter $\theta$ occurring in noncommutative
geometry suggests the existence of a very small
scale leading to wormholes that are actually
microscopic.  This would require a semi-classical
framework, as carried out in detail in reference
\cite {GL09}. Using $r/\sqrt{\theta}$ instead of
$r$ for the horizontal axis in references
\cite{NSS06, SSN09, GL09} also suggests that
small values of $r$ are intended. Using reference
\cite{RKRI} as a starting point, it is shown in
this paper that such a wormhole can be macroscopic
and that the noncommutative-geometry background
replaces the exotic matter.  Moreover, the model
presented has zero tidal forces, thereby making
it suitable for humanoid travelers.  Finally, it
is proposed that these wormholes can be made
stable to linearized radial perturbations.

\section{The solutions}\noindent
To make the wormholes suitable for humanoid travelers,
it is desirable to have a constant redshift function, so
that $\Phi'\equiv 0$, the so-called zero-tidal-force
solution proposed in reference \cite{MT88}.  Given this
condition, the Einstein field equations become
\begin{equation}\label{E:Einstein1}
    \rho(r)=\frac{1}{8\pi}\frac{b'(r)}{r^2},
\end{equation}
\begin{equation}\label{E:Einstein2}
   p_r(r)=-\frac{1}{8\pi}\frac{b(r)}{r^3},
\end{equation}
and
\begin{equation}\label{E:Einstein3}
   p_t(r)=\frac{1}{8\pi}\left(1-\frac{b(r)}{r}\right)
   \frac{b(r)-rb'(r)}{2r^2[r-b(r)]}.
\end{equation}
Here $\rho(r)$ is the energy-density, equation (\ref{E:rho}),
$p_r(r)$ is the radial pressure, and $p_t(r)$ the lateral
pressure.

Making use of equation (\ref{E:rho}), we obtain the
shape function
\begin{equation}\label{E:shape}
   b(r)=\frac{M}{\sqrt{\pi}}\left[2\sqrt{\pi}\,\text{erf}
   \left(\frac{r}{2\sqrt{\theta}}\right)-\frac{2r}
   {\sqrt{\theta}}
   \,e^{-r^2/4\theta}+C\right],
\end{equation}
where $C$ is an integration constant.  (How $M$ is related
to the mass of the wormhole will be seen at the end of
Section \ref{S:violation}.)  This solution is also given in
reference \cite{RKRI}, where it is shown that the wormhole
spacetime is asymptotically flat.

Since $C$ is an integration constant, mathematically $b(r)$
in equation (\ref{E:shape}) is a valid solution for every $C$.  The
flare-out condition, among others, is a physical requirement
that is satisfied only for a certain range of values of $C$.
To show that the wormhole is macroscopic, we use a graphical
approach, as in reference \cite{RKRI}, by assigning values to the
parameters that are not only typical but help produce useful
and revealing graphs.  Since we are using geometrized units
(with $c=G=1$), the mass $M$ takes on units of length; so
it is desirable to measure distances in meters, as in references
\cite{MT88}.  Moreover, unlike references
\cite{NSS06, SSN09, GL09}, we plot $b(r)$ against $r$,
rather than $r/\sqrt{\theta}$.  As an illustration, suppose
we choose $M=0.05$, $\theta=0.001$, and $C=3$ to obtain
$b(r)$ in Figure 1.
 \begin{figure}[htbp]
        \includegraphics[scale=.6]{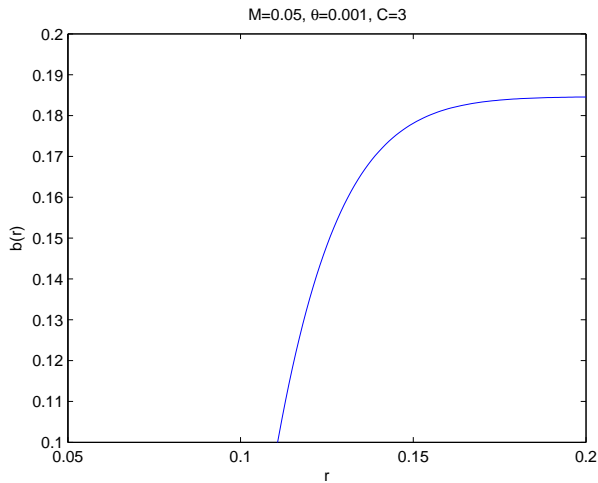}
        \caption{The shape function of the wormhole.}
   \label{fig:shape2}
\end{figure}
\begin{figure}[htbp]
        \includegraphics[scale=.6]{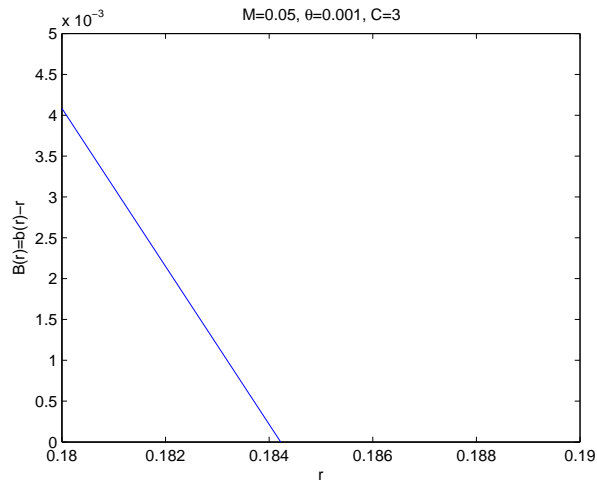}
        \caption{The throat radius is the $r$-intercept
        of $B(r)=b(r)-r$.}
   \label{fig:shape2}
\end{figure}

The throat of the wormhole is located at $r=r_0$, where $B(r)=
b(r)-r$ intersects the $r$-axis, shown in Figure 2.  The throat
radius turns out to be $r_0=0.18462$ m, which is definitely
macroscopic, a conclusion already deducible from reference
\cite{RKRI}.  (The value can be readily increased by adjusting
the parameters.)  According to Figure 2, for $r>r_0$, $B(r)<0$,
so that $b(r)-r<0$, which implies that $b(r)/r<1$ for $r>r_0$.
Also, since $B(r)$ is decreasing, we have $B'(r_0)<0$ and hence
$b'(r_0)<1$.  The flare-out condition is therefore satisfied.

\section{Violation of the weak energy condition}
\label{S:violation}\noindent
Next, recall from reference \cite{MT88} that the condition
$b'(r_0)<1$ implies that $\rho+p_r<0$, a condition that is
normally associated with exotic matter.  However, there exist
other possibilities.  For example, phantom energy is
characterized by the equation of state $p=w\rho$, $w<-1$,
which yields $\rho+p=\rho(w+1)<0$.  In the present situation,
equations (\ref{E:Einstein2}) and (\ref{E:shape}) imply that
\begin{equation}\label{E:pressure}
 p_r(r)=-\frac{1}{8\pi r^3}
   \frac{M}{\sqrt{\pi}}\left[2\sqrt{\pi}\,\text{erf}
   \left(\frac{r}{2\sqrt{\theta}}\right)-\frac{2r}{\sqrt{\theta}}
   \,e^{-r^2/4\theta}+C\right].
\end{equation}
In reference \cite{RKRI}, it is shown graphically that
$\rho+p_r<0$ near the throat, i.e.,
\begin{equation}\label{E:WEC}
   \rho+p_r=\frac{M}{(4\pi\theta)^{3/2}}e^{-r^2/4\theta}-\frac{1}
   {8\pi r^3}\frac{M}{  \sqrt{\pi}}\left[2\sqrt{\pi}\,\text{erf}
   \left(\frac{r}{2\sqrt{\theta}}\right)-\frac{2r}{\sqrt{\theta}}
   \,e^{-r^2/4\theta}+C\right]<0
\end{equation}
for certain choices of the parameters.  For the values in the above
example ($M=0.05$, $\theta=0.001$, and $C=3$), we obtain, at the
throat $r_0=0.18462$, $\rho+p_r=-0.657$.  It is important to
realize that this is a direct consequence of the
noncommutative-geometry background, rather than to some
form of ``exotic matter."  The reason is that we are not
following the strategy in reference \cite{MT88}, tailoring $b(r)$
and $\Phi(r)$ to produce a traversable wormhole, while letting
$\rho$, $p_r$, and $p_t$ dangle.  Instead, $b(r)$ and $p_r(r)$ were
obtained from $\rho(r)$ in equation (\ref{E:rho}) and the Einstein
field equations; $p_t(r)$ is derived in reference \cite{RKRI}.

Since the smearing effect is heavily influenced by the size
of $\theta$, smaller values should be considered.  (Recall that
$\theta=0.001$ was chosen mainly for producing useful graphs.)
It turns out, however, that reducing $\theta$ seems to have
little effect on the outcome with one notable exception: as
$\theta$ gets smaller, $b'(r_0)$ gets closer to zero.  This
is to be expected since $\text{lim}_{\theta
\rightarrow 0}\,b(r)=2M+MC/\sqrt{\pi}$, a constant.  The
slow flaring out, however, could have an effect on the
proper distance around the throat.  To provide a numerical
check on our example, consider the coordinate distance
given by the interval [0.18462 m, 100 m].  The
corresponding proper distance is
\[
    \int^{100}_{0.18462}\frac{dr}{\sqrt{1-b(r)/r}}
    \approx 100.61\,\text{m}.
\]
So the proper distance is not much larger than the
coordinate distance.

\emph{Remark:} The above limit, $\text{lim}_{\theta
\rightarrow 0}\,b(r)=2M+MC/\sqrt{\pi}=
\text{lim}_{r/\sqrt{\theta}\rightarrow\infty}\,
b(r)=2M+MC/\sqrt{\pi}$, shows that $2M+MC/\sqrt{\pi}=
2M_S$, the Schwarzschild limit.  So in the above example,
since $M=0.05$ and $C=3$, we obtain $M_S=0.092$ m, which
may be viewed as the mass of the wormhole relative to
a distant observer.

\section{Stability}\noindent
It is pointed out in reference \cite{pK10} that it is possible
in principle to construct a wormhole that is stable to
linearized radial perturbations.  The wormhole has to
meet the usual conditions at the throat, but when joined
to an external Schwarzschild spacetime at the junction
surface $r=a>r_0$, some additional conditions need to be
met: to be a stable wormhole, the redshift function
$\Phi=\Phi(r)$ must satisfy
the conditions
\[
   \Phi(a-)=\Phi(a+)\quad \text{and} \quad
   \Phi'(a-)=\frac{M}{a(a-2M)},
\]
referring to the left- and right-hand limits, respectively.
[In other words, $\Phi_{\text{internal}}(a)=
\Phi_{\text{external}}(a)$ and $\Phi'_{\text{internal}}(a)=
\Phi'_{\text{external}}(a)$.]  It is also shown in reference
\cite{pK10} that the shape function $b=b(r)$ must be an
increasing function of $r$ having a continuous second
derivative and that it must attain a maximum value at
$r=a$.  For such a  shape function, the stability
criterion is surprisingly simple:
\[
   b''(a)<-\frac{8M}{a^2}.
\]
The implication is that from the standpoint of theoretical
construction, this condition would not be difficult to
meet. Joining our earlier solution, namely
$\Phi\equiv \text{constant}$ (i.e., $\Phi'\equiv 0)$ and
equation (\ref{E:shape}), to this new solution will require
transitional curves that smoothly connect, respectively,
the redshift and shape functions.  The result is a stable
wormhole.

This construction does not depend on any particular
properties of the wormhole, suggesting that any wormhole
can be made stable to linearized radial perturbations.

\section{Conclusion}\noindent
It is shown in this paper that it is possible in principle to
design or construct a traversable wormhole with zero tidal
forces in a noncommutative-geometry setting.  Not only is
the wormhole spacetime macroscopic and asymptotically flat,
the violation of the weak energy condition is due to the
noncommutative geometry, rather than to ``exotic matter."
So if string theory is correct, then the laws of physics
seem to allow macroscopic traversable wormholes with zero tidal
forces.

It is also proposed in this paper that any wormhole of the
Morris-Thorne type can be modified to become stable to
linearized radial perturbations.

\end{document}